\documentclass[11pt]{article}

\usepackage{epsfig}
\usepackage{latexsym}
\usepackage{amssymb}

\title{{\bf Fluxbranes from p-branes}}
\author{{\bf P. M. Saffin}\thanks{email: p.m.saffin@durham.ac.uk}
\\ Centre for Particle Theory, Department of Mathematical Sciences, \\
University of Durham, South Road, Durham DH1 3LE, 
United Kingdom.\\
\\ \\ 
}

\date{\today}

\begin{document}

\maketitle
\begin{abstract}
It is shown how
magnetic fluxbrane, Fp-brane, solutions 
are related to electric
black p-branes by analytic continuation. Viewing 
the transverse space of branes as a warped cone, one finds that
the cone base of the p-brane
becomes the world-volume of the F(D-p-3)-brane and the
world volume of the p-brane becomes the cone base of
the F(D-p-3)-brane.
An explicit
example of the correspondence is given for a 2-brane  and
F6-brane of 11D supergravity.
\end{abstract}

\section{Introduction}

A long standing solution to Einstein-Maxwell theory is the Melvin
Universe \cite{melvin64},
describing how parallel lines of magnetic (electric)
flux self-gravitates to form a classically stable flux-tube
\cite{thorne65}.
The solution was later extended to dilaton-Einstein-Maxwell
theory in arbitrary dimensions
in order to give it a place in supergravity theories
\cite{gibbons88}. 
It was later discovered that the dilaton-Melvin solution has
a rather distinguished heritage, as it can be derived from pure
gravity in one dimension higher by compactification of Minkowski
space \cite{dowker94}. This aspect of the solution then leads
it to being an exact sigma model background for string theory
\cite{tseytlin95}. Further interest in the Melvin solution comes
from duality arguments; it is
believed that a Melvin background with field strength $B$
in type 0A and a Melvin
background with $B'$ of type IIA are equivalent, for specific
$B$, $B'$ \cite{0AIIA}.
With these aspects in mind an attempt was made to generalize the
Melvin solution from a fluxbrane due to a 2-form field strength to
a fluxbrane for an arbitrary rank field strength \cite{saffin01},
see also \cite{galtsov99}.
The near core and asymptotic behaviour of a
class of fluxbranes was studied in \cite{gutperle01}.
The physical picture of these fluxbranes comes from giving a
p-brane and anti-p-brane infinite separation, the flux from the
p-brane is transmitted to the anti-p-brane along F(p+1)-branes.

There is yet another aspect to Fp-branes which warrants
interest, namely the dielectric brane effect \cite{myers99}. When
a number of D0-branes are placed in a constant electric
background 4-form field strength one finds that the 0-branes expand
to form a non-commutative two-sphere. In terms of supergravity this
picture cannot be entirely correct as there is no such thing as
a constant form field strength, the form self-gravitates to create
Fp-branes. A supergravity description of just such a system has been
found in \cite{emparan01}, \cite{herdeiro01}..

Here we look at Fp-branes from a perspective prompted by \cite{gibbons01}.
In \cite{gibbons01} it was shown how one can start
with the Reissner-Nordstr\"{o}m (RN) 
metric of an electrically charged
black hole and, after a limiting procedure and analytic
continuation, arrive at the Melvin solution.
We aim to place this result in the more general setting of generating
magnetic F(D-p-3)-branes from electrically charged p-branes.
In fact such a relation was already hinted at in \cite{gibbons88} where
0-branes (black holes) and (D-3)-fluxbranes were studied together because
of the similarity of the ans\"{a}tze.
(See also \cite{gibbons87} where a flux tube solution is related to the
analytic continuation of a cosmological model.)
 Although we only explicitly work
with electric p-branes and magnetic Fp-branes, electric/magnetic duality
means there is a similar relation between 
magnetic p-branes and electric Fp-branes

The outline of the paper is to start by covering the RN-Melvin connection,
to make the procedure clear, and with this in mind the general 
correspondence will be found. To see how this works in practise we take
the non-trivial example of turning an electric 2-brane of D=11 
supergravity
into a magnetic F6 brane. Conclusions are then drawn at the end.

\setcounter{equation}{0}
\section{Melvin from Reissner-Nordstr\"{o}m}

The action that we shall be studying is that of gravity plus n-form
in $D$ dimensions,
\begin{eqnarray}
S&=&\int\left[R *1-\frac{1}{2}{\rm F}_{(n)}\wedge *{\rm F}_{(n)}\right].
\end{eqnarray}
which gives the following equations of motion,
\begin{eqnarray}
{\cal R}_{MN}&=&\frac{1}{2(n-1)!}\left[F_{M...}F_N^{\;\;\;...}
                 -\frac{n-1}{n(D-2)}F^2g_{MN}\right]\\
{\rm d}*F&=&0
\end{eqnarray}
for the Ricci tensor ${\cal R}_{MN}$ and $F_{(n)}$.
In this paper we shall not include the dilaton as it does not add
anything to the discussion.

The starting point for the derivation is the RN metric of four dimensional
Einstein-Maxwell theory,
\begin{eqnarray}
{\rm ds}^2&=&-V{\rm d}t^2+V^{-1}{\rm d}r^2+r^2{\rm d}\Omega_{(2)}^2\\
\label{Vfunc}
V&=&\frac{1}{r^2}(r-r_-)(r-r_+)\\
F_{(2)}&=&\frac{2\sqrt{r_- r_+}}{r^2}{\rm d}t\wedge{\rm d}r
\end{eqnarray}
where ${\rm d}\Omega_{(2)}^2$ is the usual spherical volume element.
This is a metric describing an electrically charged black hole, and
if $r_\pm$ are positive it has two horizons, at $r=r_\pm$.

We start with a coordinate change and a redefinition of parameters,
\begin{eqnarray}
\label{radcoord}
r&=&r_-+\frac{1}{4}B^2r_-\rho^2,\\
\label{horizons}
r_+&=&-r_-^3B^2,
\end{eqnarray}
noting that now there is no horizon at $r=|r_+|$, at the expense if an imaginary
electric charge.

A limiting procedure is now performed to flatten the spherical volume
element,
using stereographic coordinates for the sphere makes this explicit,
\begin{eqnarray}
{\rm d}\Omega_{(2)}^2&=&\frac{4}{(1+\xi\bar{\xi})^2}{\rm d}\xi{\rm d}\bar{\xi}\\
r_-&\rightarrow& r_-/\lambda,\;\;
\xi\rightarrow \lambda(x+iz),\;\;
t\rightarrow \lambda t,\\
\lambda&\rightarrow& 0.
\end{eqnarray}
This is a way of getting the equivalent of the RN solution with the
spherical volume element replaced by a Ricci flat metric. 
Such a replacement means that we find a spacetime which is not asymptotically
flat, however, neither is the Melvin solution which is what we are aiming for.

Now define,
\begin{eqnarray}
{\rm t}=\frac{2i}{B^2r_-}\phi,\;\;
x=i\frac{T}{2},\;\;
z=\frac{Z}{2},\;\;
\end{eqnarray}

With this done we find the metric and field strength,
\begin{eqnarray}
{\rm ds}^2&=&\Lambda(\rho)^2\left(-{\rm d}T^2+{\rm d}Z^2\right)
             +\Lambda(\rho)^2{\rm d}\rho^2
             +\rho^2\Lambda(\rho)^{-2}{\rm d}\phi^2\\
\Lambda(\rho)&=&1+\frac{B^2}{4}\rho^2\\
F_{(2)}&=&2B\rho\Lambda(\rho)^{-2}{\rm d}\rho\wedge{\rm d}\phi
\end{eqnarray}
which is just the magnetic Melvin solution. We note here that the inner horizon,
$r_-$, of the black hole has become the origin of the Melvin solution, 
(\ref{radcoord}). The regularity of this inner black hole horizon 
now translates
to the regularity of the core of the flux-tube, a point made in 
\cite{gibbons88}.
It is also worth noting that the two horizons had to have different 
values, (\ref{horizons}), this
therefore precludes the use of extremal branes. To be more explicit,
suppose we started with the extremal case of \mbox{$r_+=r_-=r_H$} then
\begin{eqnarray}
F_{\rm extreme}&=&2\frac{r_H}{r^2}{\rm d}t\wedge{\rm d}r
\end{eqnarray}
and the analytic continuation of the world volume then takes 
\mbox{$t\rightarrow i\tau$}. To keep the form field real we then take
\mbox{$r_H\rightarrow iR_H$} but this makes the metric
function $V(r)$ (\ref{Vfunc})
imaginary. This is the typical situation, starting with
the extreme p-branes leads to fluxbranes with imaginary coefficients,
either in the metric or form field, or both.

The important observation to make is that the world-volume of the black hole,
$dt$, has become the angular coordinate of the flux-tube and the angular
coordinates of the black hole, d$\Omega_{(2)}$, have become the world volume
of the flux-tube.

We shall see that this type of relation holds in the general case. Recalling
that the metric of a cone with base metric d$\omega^2$ is 
\mbox{${\rm d}r^2+r^2{\rm d}\omega^2$}, we can view the space transverse to
the world-volume of a brane as a warped cone. The correspondence is then that
the cone bases get analytically continued to become world-volumes and 
world-volumes are analytically continued into cone bases.
As a side remark, we note that had we not flattened the
${\rm d}\Omega_{(2)}^2$ we would have analytically continued this to a
deSitter metric, giving a flux-tube with deSitter world-volume. 

\setcounter{equation}{0}
\section{General correspondence}

The relation between magnetic F-branes and electric p-branes is made
clearer by writing down the ansatz for each of them. We start with the
electric p-brane from an $n$-form field strength in $D$ dimensions; p$=n-2$,
$d=n-1$, $\tilde{d}=D-d-2$,
\begin{eqnarray}
{\rm dS}^2({\rm p-brane})&=& \exp\left(2A(r)\right)\bar{{\rm ds}}^2_{(d)}
                     +\exp\left(2B(r)\right){\rm d}r^2
                     +\exp\left(2C(r)\right){\rm ds}^2_{(\tilde{d}+1)}\\
{\rm F}_{d+1}&\propto&\exp\left(dA+B-(\tilde{d}+1)C\right)
                \eta_{(d)}\wedge{\rm d}r.
\end{eqnarray}
While an $n$-form field strength gives rise to the following 
magnetic Fp-brane in
$D$ dimensions \cite{saffin01}; \mbox{p$=D-n-1$}, $m=n-1$, $l=D-m-2$,
\begin{eqnarray}
{\rm dS}^2({\rm Fp-brane})&=& \exp\left(2a(r)\right)\bar{{\rm ds}}^2_{(l+1)}
                     +\exp\left(2b(r)\right){\rm d}r^2
                     +\exp\left(2c(r)\right){\rm ds}^2_{(m)}\\
{\rm F}_{m+1}&\propto&\exp\left(-(l+1)a+b+mc\right)
                      \eta_{(m)}\wedge{\rm d}r.
\end{eqnarray}
The volume elements $\bar{{\rm ds}}^2$ have Lorentzian signature and
describe the world volumes of the branes. The transverse space is made
up of a radial coordinate $r$ and Euclidean signature metrics ds$^2$,
so ds$^2$ are the cone bases for the transverse space.
The similarity of the two systems should now be clear, with us being
able to move from the p-brane description to the Fp-brane by 
changing $d\rightarrow m$, $\tilde{d}\rightarrow l$, 
$A(r)\rightarrow c(r)$, $B(r)\rightarrow b(r)$, $C(r)\rightarrow a(r)$.
The world-volume metric, $\bar{\rm ds}^2_{(d)}$, gets analytically
continued
to become the cone base metric, ${\rm ds}^2_{(m)}$, and the cone base
metric, ${\rm ds}^2_{(\tilde{d}+1)}$, 
gets continued to become the world-volume
metric, $\bar{\rm ds}^2_{(l+1)}$.

Given this correspondence we should ask whether the well known p-brane
solutions can uncover new fluxbrane solutions. The Fp-branes of most
interest would be those where the cone base was a sphere with its round
metric. The correspondence says that these are equivalent to p-branes
with deSitter world-volume, a system which has not yet been solved.

We can however use the correspondence to understand some features of 
\cite{saffin01}. That paper described Fp-branes where the world-volume
and cone base metrics were Einstein metrics, independent of the radial
coordinate. These then should match with black p-branes whose world volume
and cone base are also Einstein and independent of the radial distance.
Given this it is clear that the `standard' black branes of Duff
{\it et al} \cite{duff96} will not analytically continue to these Fp-branes
as there the world volume has a radial dependence.
There is a class of black brane
solutions however where the
world volume is flat and has a boost symmetry \cite{gregory96}, these should
therefore be continued to give Fp branes where the cone base is flat.
These black brane solutions then explain some properties of the Fp-branes
found in \cite{saffin01}. The Fp-branes with deSitter world volume and
flat cone base were found to be asymptotically flat, this now translates
to the asymptotic flatness of the black brane solutions with Minkowski
world volume. Similarly, the non-asymptotic flatness of Fp-branes with
Ricci flat world volume gets translated to the same property of
black branes with a Ricci flat cone base.

\setcounter{equation}{0}
\section{Explicit example; $D=11$ F6-brane from the 2-brane.}

We now give an explicit, non trivial example of how this correspondence
relates the black branes of \cite{gregory96} to the Fp-branes of
\cite{saffin01}. The example we choose is motivated by 11D supergravity,
starting with
an electric 2-brane we generate a magnetic F6-brane as follows.

Taking note of the different convention for the form normalization we
start with the electric 2-brane solution of \cite{gregory96},
\begin{eqnarray}
{\rm ds}^2&=&
\Lambda^{-\frac{2}{3}}{\rm d}\underline{x}_{(3)}^2\\
\nonumber
&+&
\Lambda^{\frac{1}{3}}
        \left(\frac{r^6-r_+^6}{r^6-r_-^6}\right)^{-\frac{5}{6}}
        \left(r^6-r_+^6\right)^{-\frac{5}{3}}
       r^{10}{\rm d}r^2\\
\nonumber
&+&
\Lambda^{\frac{1}{3}}
        \left(\frac{r^6-r_+^6}{r^6-r_-^6}\right)^{\frac{1}{6}}
        \left(r^6-r_+^6\right)^{\frac{1}{3}}
       {\rm d}\Omega^2_{(7)},\\
F_{(4)}&=&2Q*\eta_{(7)},\\
Q^2&=&9r_-^6\left[\sqrt{\frac{7}{3}}\left(r_+^6-r_-^6\right)+r_-^6\right],\\
\Lambda&=&\left\{\left(1+\frac{r_-^6}{\sqrt{\frac{7}{3}}(r_+^6-r_-^6)}\right)
      \left(\frac{r^6-r_+^6}{r^6-r_-^6}\right)^{-\frac{1}{2}\sqrt{\frac{7}{3}}}
      -\frac{r_-^6}{\sqrt{\frac{7}{3}}(r_+^6-r_-^6)}
      \left(\frac{r^6-r_+^6}{r^6-r_-^6}\right)^{\frac{1}{2}\sqrt{\frac{7}{3}}}
\right\}
\end{eqnarray}
With $\eta_{(7)}$ being the volume form for the round 7-sphere.
Following in analogy with the RN-Melvin case we take the parameter describing
the outer horizon, $r_+$, to be negative and we define a new radial coordinate,
$R$,
\begin{eqnarray}
r_+^6=-B^{2c_0},\;\;
\frac{r^6-r_+^6}{r^6-r_-^6}=\left(\frac{\Gamma R}{B}\right)^{2c_0}.
\end{eqnarray}
Two radial scales, $R_0$ and $R_1$, are introduced by the relations
\begin{eqnarray}
\left(\frac{B}{\Gamma R_1}\right)^{2c_0\sqrt{\frac{7}{3}}}
&=&\frac{\sqrt{\frac{7}{3}}\left(r_+^6-r_-^6\right)+r_-^6}{r_-^6}\\
R_0&=&\frac{B}{\Gamma}
\end{eqnarray}
and we introduce the constant $\kappa$ by,
\begin{eqnarray}
\left(B^{2c_0}+r_-^6\right)\left[ \left(\frac{R_0}{R_1}\right)^{c_0\sqrt{\frac{7}{3}}}
                   +\left(\frac{R_0}{R_1}\right)^{-c_0\sqrt{\frac{7}{3}}}\right]^{-1}
&=&\frac{\kappa}{6\sqrt{\frac{7}{3}}}
\end{eqnarray}
from which we find,
\begin{eqnarray}
{\rm ds}^2&=&\left(\frac{c_0}{3}\right)^{\frac{2}{3}}
             \left(\frac{\kappa}{6\sqrt{\frac{7}{3}}}\right)^{\frac{2}{3}}
             \tilde{\Lambda}^{-\frac{2}{3}}
          \left(\frac{3}{c_0}\right)^{\frac{2}{3}}
          \left(B^{2c_0}+r_-^6\right)^{\frac{2}{3}}{\rm d}\underline{x}_{(3)}^2\\
\nonumber
&+&\frac{c_0^2}{9}\left(\frac{\kappa}{6\sqrt{\frac{7}{3}}}\right)^{\frac{1}{3}}
      \tilde{\Lambda}^{\frac{1}{3}}
      \left(  \left(\frac{R}{R_0}\right)^{-c_0}
             -\left(\frac{R}{R_0}\right)^{c_0}
      \right)^{-\frac{7}{3}}\frac{{\rm d}R^2}{R^2}\\
\nonumber
&+&\left(\frac{\kappa}{6\sqrt{\frac{7}{3}}}\right)^{\frac{1}{3}}
     \tilde{\Lambda}^{\frac{1}{3}}
      \left(  \left(\frac{R}{R_0}\right)^{-c_0}
             -\left(\frac{R}{R_0}\right)^{c_0}
      \right)^{-\frac{1}{3}}{\rm d}\Omega_{(7)}^2\\
F&=&2\kappa\left(\frac{c_0}{3}\right)^2
     \left(\frac{\kappa}{6\sqrt{\frac{7}{3}}}\right)^{-2}
    \tilde{\Lambda}^{-2}
    \left(\frac{3}{c_0}\right)
    \left(B^{2c_0}+r_-^6\right)
    i{\rm d}x^0\wedge{\rm d}x^1\wedge{\rm d}x^2\wedge\frac{{\rm d}R}{R}\\
\tilde{\Lambda}&=&\left\{ \left(\frac{R}{R_1}\right)^{c_0\sqrt{\frac{7}{3}}}
                    +\left(\frac{R}{R_1}\right)^{-c_0\sqrt{\frac{7}{3}}}
             \right\}
\end{eqnarray}
The final step is a rescaling of the world-volume coordinates and the analytic
continuation,
\begin{eqnarray}
\left(\frac{3}{c_0}\right)^{\frac{1}{3}}
\left(B^{2c_0}+r_-^6\right)^{\frac{1}{3}}\underline{x}=\underline{Y}\\
(iY^0,Y^1,Y^2)\rightarrow(X^9,X^{10},X^{11}),\\
{\rm d}\Omega_{(7)}\rightarrow{\rm d}\omega_{(7)},
\end{eqnarray}
where ${\rm d}\omega_{(7)}$ is the deSitter volume element, the analytic
continuation of the round sphere metric.

What we have ended up with is an F6-brane with deSitter world volume and
a transverse space which has a flat cone base, as such we may compare
it to the metric and field strength found in
\cite{saffin01}, section 6 with $m=3$, $l=6$, $\Lambda_{(L)}=1$,
which should describe such a fluxbrane,
\begin{eqnarray}
{\rm ds}^2&=&\exp\left(2a(\xi)\right){\rm d}\omega^2_{(7)}
             +\exp\left(2(A(\xi)+a(\xi)\right)
             +\exp\left(2c(\xi)\right){\rm d}\underline{X}^2_{(3)}\\
F&=&\kappa\exp\left(6c(\xi)\right)
    {\rm d}x^9\wedge{\rm d}x^{10}\wedge{\rm d}x^{11}\wedge{\rm d}\xi\\
A(\xi)&=&-\ln\left\{-\frac{6}{c_0}\sinh\left[c_0(\xi-\xi_0)\right]\right\}\\
c(\xi)&=&-\frac{1}{3}\ln\left\{-\frac{\kappa}{c_0\sqrt{\frac{7}{3}}}
           \cosh\left[c_0\sqrt{\frac{7}{3}}(\xi-\xi_1)\right]\right\}\\
a(\xi)&=&\frac{1}{6}(A(\xi)-3c(\xi)),\;\;
b(\xi)=A(\xi)+c(\xi).
\end{eqnarray}
After a coordinate change, \mbox{$\xi=\ln (R)$} and redefinition
of constants,
\mbox{$\xi_0=\ln (R_0)$}, \mbox{$\xi_1=\ln (R_1)$} we find precise agreement
between the metric and form field derived from the black p-brane.

\setcounter{equation}{0}
\section{Conclusions.}

The well known objects of supergravity theory, p-branes, have been related
to their less studied cousins, Fp-branes. We found that
the argument of Gibbons and Herdeiro \cite{gibbons01},
which relates the Reissner-Nordstr\"{o}m black hole to the Melvin
flux-tube, can be extended to a more general relation between 
electric black p-branes
and magnetic Fp-branes. The end result being 
that the transverse space cone base (world volume) 
of a p-brane gets analytically continued to the world volume 
(transverse space cone base) of an F(D-p-3) brane.
In particular, the interesting case of an Fp-brane with
a round sphere for the cone base relates to the problem
of a black brane with deSitter world volume.

\section{Acknowledgements}
I would like to thank Dom Brecher, Bert Janssen,
Simon Ross and Douglas Smith for useful conversations
and suggestions, 
and PPARC for financial support.


\end{document}